\documentclass[twocolumn,english,aps,pra,longbibliography,superscriptaddress,amsmath,amssymb,floatfix]{revtex4-2}
\usepackage[utf8]{inputenc}
\usepackage{microtype}
\usepackage{graphicx}
\usepackage{amsmath}
\usepackage{amssymb}
\usepackage{mathtools}
\usepackage[colorlinks=true,urlcolor=blue,citecolor=blue,linkcolor=blue]{hyperref}
\usepackage{natbib}
\usepackage{physics}
\usepackage{cleveref}
\usepackage{sidecap}
\usepackage{makecell}
\usepackage[linesnumbered,lined,commentsnumbered]{algorithm2e}
\Crefname{algocf}{Algorithm}{Algorithms}
\usepackage{layouts}
\usepackage{dcolumn}
\usepackage{bm}

\usepackage{dsfont}
\usepackage{comment}
\usepackage{float}

\usepackage{pifont}     
\usepackage{xcolor}     
       
\usepackage{tabularx}
\usepackage{booktabs}

\begin{document}
\preprint{APS/123-QED}

\title{Amplification of bosonic interactions through squeezing in the presence of decoherence}

\date{\today}

\author{Ankit Tiwari}
\affiliation{School of Electrical, Computer, and Energy Engineering, Arizona State University, Tempe, Arizona 85287, USA}
\author{Cecilia Cormick}
\affiliation{Instituto de F\'isica de la Facultad de Ingeniería, Universidad de la República, Julio Herrera y Reissig 565, 11300 Montevideo, Uruguay}
\affiliation{Instituto de F\'isica Enrique Gaviola, CONICET and Universidad Nacional de C\'ordoba,
Ciudad Universitaria, X5016LAE, C\'ordoba, Argentina}
\author{Christian Arenz}
\affiliation{School of Electrical, Computer, and Energy Engineering, Arizona State University, Tempe, Arizona 85287, USA}

\begin{abstract}
We consider the amplification of bosonic interactions through parametric control that implements squeezing along orthogonal quadratures.  We show that bosonic interactions described by certain classes of quadratic and quartic Hamiltonians can be enhanced in this way while simultaneously overcoming noise and decoherence. In general, the amplification method enhances both desired and undesired interactions present in the system. Depending on the case, however, detrimental processes can be less amplified than the desired couplings. We leverage this observation to improve the fidelity for preparing Bell-type entangled states between two bosonic modes in the presence of noise and losses. We also investigate noise models for which the protocol either fails or partially achieves a noise-tolerant state preparation speedup. Our work facilitates faster preparation of complex quantum states and implementation of entangling gates in the presence of decoherence mechanisms.
\end{abstract}

\maketitle
\section{Introduction}
Squeezing is a key resource for applications in quantum information science. For example, it plays a critical role in quantum sensing when the reduced uncertainty in one quadrature is used to measure a signal, or phase shift, more precisely \cite{lawriequantumsensing, PhysRevD.23.1693}. More generally, the amplification of quantum processes through squeezing can be used to enhance weak forces and desired interactions \cite{PhysRevLett.120.093601, PhysRevB.104.L220503,PhysRevLett.125.153602, PhysRevX.7.021041, PhysRevLett.122.030501, PhysRevLett.125.203601, PhysRevA.100.043417}, including optomechanical \cite{PhysRevLett.114.093602, enhancedaaclark} and Jaynes-Cummings couplings between quantum harmonic oscillators and spin degrees of freedom \cite{burd2023experimental, PhysRevA.107.032425}, as well as interactions appearing in superconducting circuits \cite{PhysRevLett.120.093602,PRXQuantum.5.020306, cai2025quantum, Xiong_2018}. The potential of this kind of parametric driving to speed up a desired evolution is not only a theoretical idea: proof-of-principle demonstrations have already been carried out with trapped ions \cite{burd2023experimental, scBurdQuantumampbosonic, doi:10.1126/science.aaw2884, PhysRevA.107.032425}, superconducting qubits \cite{PRXQuantum.5.020306} and photonic platforms~\cite{3bdr-njhy}.

\begin{figure}[ht!]
 \centering
\includegraphics[scale = 0.75]{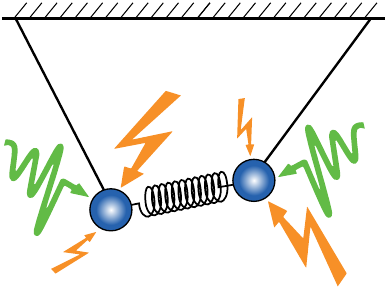}
\caption{Schematic illustration of parametric control (green) used to enhance the interaction strength between two quantum harmonic oscillators, here represented as pendulums coupled via a spring, in the presence of noise and decoherence (orange arrows). We study high-frequency periodic parametric controls implementing squeezing sequences that can amplify bosonic interactions more than detrimental processes, thereby outperforming the effect of noise and decoherence.}
\label{fig: coupled oscillators}
\end{figure}

The enhancement of unitary evolution through squeezing often comes with a cost, since undesired dynamics can simultaneously be amplified too. It was recently shown theoretically in \cite{Tiwari:25} that photon loss can be outperformed in a system of two bosonic modes coupled via an amplified cross-Kerr interaction. The desired unitary dynamics are boosted by a sequence in which one rapidly alternates between squeezing along orthogonal quadratures, a protocol referred to as \emph{Hamiltonian amplification} \cite{arenz2020amplification}. In this work, we generalize this result to a broader class of interactions for different models of noise and decoherence, a scenario that is sketched in Fig.~\ref{fig: coupled oscillators}. 

Specifically, we consider interactions that are quadratic ({\it i.e.} beamsplitter-type) and quartic ({\it i.e.} cross-Kerr) in the bosonic annihilation and creation operators. The undesired processes are modeled as random displacements and local decoherence channels described by Lindblad superoperators. We design and investigate squeezing protocols that amplify both desired and undesired dynamics, but with different rates. We show that this difference can be leveraged to comparatively enhance a given target process, taking as an illustrative case the preparation of an entangled state. Ultimately, this provides an example of how the fast parametric control implementing the squeezing sequences can open a new pathway for controlling open quantum systems, by making the desired dynamics faster than the detrimental processes.  

The manuscript is organized as follows. In Sec. \ref{sec:HAandGeneralization} we start by introducing the bosonic interactions we study in this work - beamsplitter and cross-Kerr couplings. We go on to describe the simplest Hamiltonian amplification protocol, followed by generalizing it to the two-mode interactions considered. In Sec. \ref{sec:AmpInPresenceofNoise} we study Hamiltonian amplification in the presence of noise, introducing random displacements and decoherence described by Lindblad operators. We focus on improving the fidelity for preparing Bell-type entangled states under the action of these models of uncontrolled environments. Finally, we discuss the conditions under which the amplification of the desired interactions can be achieved while simultaneously overcoming the associated noise and decoherence processes. In Sec. ~\ref{sec:platforms} we describe different experimental platforms in which our protocols could be applicable, and provide suitable references for the interested reader. Our findings are summarized in Sec.~\ref{sec:conclusion}. Technical details and further examples are provided in Appendices \ref{sec:appendix1} to \ref{sec:appendix3}.

\section{Hamiltonian amplification of bosonic interactions}\label{sec:HAandGeneralization}

\subsection{Two-mode interactions}

Our system of interest is formed by two bosonic modes, with annihilation (creation) operators $a$ ($  a^{\dagger}$) and $b$ ($b^{\dagger}$). The frequencies of the corresponding quantum harmonic oscillators are referred to as $\omega_a$ and $\omega_b$, respectively, and we set $\hbar=1$. In this work, we consider two kinds of couplings between the modes. The first one corresponds to a beamsplitter-type interaction described by the Hamiltonian  
\begin{equation}\label{eq:beamsplitter}
H_{\text{bs}} = g (ab^{\dagger} + a^{\dagger}b),    
\end{equation}
where $g$ is the coupling strength. This induces an exchange of excitations between modes $a$ and $b$ and is analogous to the action of an optical beamsplitter \cite{Walls-Milburn}. 

The second case we consider is a cross-Kerr interaction given by the Hamiltonian
\begin{equation}\label{eq:cross-Kerr}
H_{\text{cK}} =  \chi \, a^{\dagger}a\, b^{\dagger}b,  
\end{equation}
where $\chi$ is the strength of the cross-Kerr effect. This coupling shifts the frequency of each oscillator conditioned on the occupation number of the other mode \cite{PhysRevLett.62.2124,RALPH2010209}. This conditional phase shift can be used to implement a controlled-phase gate in photonic systems to enable universal photonic-based quantum computations \cite{PhysRevLett.62.2124}.  

For simplicity, in the following we will work in the interaction picture with respect to the uncoupled evolution so that the coupling Hamiltonians will be the only generators of the evolution. When considering beamsplitter interactions, we will assume that both oscillator frequencies are equal so that the energy exchange is resonant and the interaction picture does not introduce any time dependence in the Hamiltonian. We note, however, that this resonance condition, satisfied naturally in scenarios like \cite{Harlander_antenna}, is not strictly required by the protocol. Instead, Hamiltonian~\eqref{eq:beamsplitter} could result from the use of an additional time-dependent potential facilitating the exchange of excitations \cite{Hou_NaturePhys_2024}. 

Either coupling, \eqref{eq:beamsplitter} or \eqref{eq:cross-Kerr}, can be used to generate Bell-type states. In the case of the beamsplitter interaction~\eqref{eq:beamsplitter}, a Bell-type state can be generated by first preparing the separable state $\ket{01}$ and then letting it evolve under $U(t) = \exp(-iH_{\text{bs}}t)$
 for a time $t_{\Phi} = \frac{\pi}{4g}$, to  obtain
\begin{equation}\label{eq:entangledstate1}
 \ket{\Phi} = \frac{1}{\sqrt{2}} ( \ket{01} + i \ket{10}),    
\end{equation}
where $\ket{n}$ $(n \in \mathbb{N}_0)$ denotes the eigenstates of the number operators $a^{\dagger}a$ and $b^{\dagger}b$. We note that the operation we consider is often referred to as an $\sqrt{i\,\rm{SWAP}}$-gate, since by applying it twice one exchanges the bosonic operators of the two modes up to a $\pi/2$ phase.  

Similarly, one can generate a Bell-type state under the cross-Kerr evolution generated by \eqref{eq:cross-Kerr} by first preparing the separable state $\ket{++}$ and then letting it evolve under $U(t) = \exp(-i H_{\text{cK}}t)$ for a time $t_{\Psi} = \frac{\pi}{\chi}$ so that the final state is of the form:
\begin{equation}\label{eq:entangledstate2}
 \ket{\Psi} = \frac{1}{\sqrt{2}} ( \ket{0\, +} +  \ket{1\, -}),
\end{equation}
where we use the notation $\ket{\pm} = \frac{1}{\sqrt{2}} (\ket{0} \pm \ket{1} )$.

\subsection{Basic Hamiltonian amplification scheme}

The interaction strengths in \eqref{eq:beamsplitter} and \eqref{eq:cross-Kerr} can be enhanced through a local squeezing protocol called Hamiltonian amplification \cite{arenz2020amplification}. The scheme can be most easily understood in a bang-bang implementation \cite{lidar2013quantum} for a single mode where sudden pulses instantaneously implement squeezing transformations along different quadratures. Considering a quantum harmonic oscillator with frequency $\omega$, the method involves the application of squeezing operations 
$S_{a,\theta}(r) = \exp\left[\frac{r}{2}\left(  a^{2}e^{-i\theta}-  a^{\dagger 2}e^{i\theta}\right)\right]$, where $r$ is the squeezing strength and $\theta \in \{0, \pi\}$ describes the quadrature along which the mode is squeezed. These operations are alternatively applied to the system after short periods of free evolution, resulting in a total (Trotterized) evolution given by:
\begin{equation}\label{eq: HA sequence cosh(2r)}
 U_{N}(t) = \left[S^{\dagger}_{a,0}U(\Delta t )S_{a,0} S^{ \dagger}_{a,\pi}U(\Delta t )S_{a,\pi}\right]^{N}.
\end{equation}
Here, $\Delta t = \frac{t}{2N}$, $N$ is the number of times the sequence is repeated, and $U(\Delta t ) = \exp(-iH_{0}t)$ with $H_0$ the relevant Hamiltonian, in this case corresponding to the quantum harmonic oscillator described by $H_{0}=\omega a^{\dagger}a$. In the limit $N\to \infty$, referred to as the Trotter limit, the resulting sequence is exactly equivalent to $U_\infty(t) = \exp(-i H_\lambda t)$ with $H_\lambda = \lambda H_0$ an amplified Hamiltonian. In particular, in this case, one has $\lambda = \cosh(2r)$ so that the speed-up is directly related to the amount of squeezing.

As discussed in \cite{arenz2020amplification}, for finite values of $N$ Trotter errors are introduced so that, in general, there is a trade-off between the number of steps and fidelity. However, it is important to note that both the Trotter error and the experimental challenges of a bang-bang protocol can be significantly reduced by applying a continuous version of the above scheme. In this case, instead of a sudden squeezing operation, one applies a squeezing Hamiltonian which changes sign in a fast but smooth manner. By appropriately choosing the time dependence of this Hamiltonian, for instance with a sinusoidal function, as discussed in Sec. \ref{subsec:smooth}, one can eliminate the second-order Trotter error to speed up the convergence to the amplified evolution. 

\subsection{Hamiltonian amplification for two-mode interactions} \label{sec:Generalization}

The amplification scheme described above can be applied in a straightforward manner to the case of beamsplitter interactions. In this scenario, one simply needs to take $H_{0} = H_{\text{bs}}$ as given in~\eqref{eq:beamsplitter}, and replace $S_{a,\theta}(r)$ by a product of single-mode squeezing transformations: 
\begin{equation}
    S_{\theta_a,\theta_b} = S_{a,\theta_a} (r) \, S_{b,\theta_b}(r).
\end{equation}
The interaction strength can then be amplified through the sequence
\begin{equation}
\begin{aligned}\label{eq:two mode HA sequence beamsplitter}
U^{\text{bs}}_{N}(t) = \Bigl[&S_{0,0}^{\dagger}U(\Delta t)S_{0,0}S_{\pi,\pi}^{\dagger}U(\Delta t)S_{\pi,\pi}\Bigr]^{N}.
\end{aligned}
\end{equation}

Then, similarly to the single-mode Hamiltonian amplification protocol, one obtains in the Trotter limit $N\to\infty$ an effective amplified Hamiltonian $H_{\lambda_2} = \lambda_2 H_{\text{bs}}$, where $\lambda_2 = \cosh(2r)$. Here, the subindex ``2'' is used to indicate that the squeezing is applied to both modes. Employing this amplified Hamiltonian results in a speed-up of the preparation of the entangled state $\ket{\Phi}$ such that the interaction time is modified to $t_{\Phi}/{\lambda_2}$. More generally, if different squeezing strengths are employed in the operations $S_{a,\theta}(r_a)$ and $S_{b,\theta}(r_b)$, then one obtains for the beamsplitter interaction a corrected amplification factor of $\lambda_{2} = \cosh(r_a)\cosh(r_b) + \sinh(r_a)\sinh(r_b)$. For simplicity, in this work we will assume that the squeezing strengths are equal for both modes.

A generalization of the same idea can also be used to enhance the cross-Kerr interaction, as discussed in~\cite{Tiwari:25}, to speed up the implementation of a controlled-phase gate.  
Indeed, for $H_{0} = H_{\text{cK}}$ as given in~\eqref{eq:cross-Kerr} an even higher amplification factor $\lambda_2 = \cosh^2(2r)$ can be achieved by employing squeezing operations that are different for the two modes, via the sequence
\begin{equation}
\begin{aligned}\label{eq:two mode HA sequence cross-Kerr}
U^{\text{cK}}_{N}(t) = R(t)\Bigl[&S_{0,0}^{\dagger}U(\Delta t)S_{0,0}S_{\pi,0}^{\dagger}U(\Delta t)S_{\pi,0}\\
&S_{0,\pi}^{\dagger}U(\Delta t)S_{0,\pi}S_{\pi,\pi}^{\dagger}U(\Delta t)S_{\pi,\pi} \Bigr]^{N} .
\end{aligned}
\end{equation}
Here, $ \Delta t = \frac{t}{4N}$ and $R(t) = \exp(i H_R t)$ is a global phase-space rotation, with $H_R = \chi \cosh(2r)\sinh^2(r) [a^{\dagger}a + b^{\dagger}b ]$. We note that some of the single-mode squeezing operations in \eqref{eq:two mode HA sequence cross-Kerr} cancel out, but this form was chosen for compactness and readability.

For $N \to \infty$ this protocol reduces the interaction time for the preparation of the entangled state \eqref{eq:entangledstate2} by a factor $\cosh^2(2r)$. In the more general case when different squeezing strengths are employed in the operations $S_{a,\theta_1}(r_a)$ and $S_{b,\theta_2}(r_b)$, the cross-Kerr amplification factor takes the form $\lambda_{2} = \cosh(2r_a)\cosh(2r_b)$ and the rotation $R$ is a product of a phase-space rotation for each mode, where the two rotation angles are no longer equal. 

\subsection{Hamiltonian amplification through smooth parametric control}

\label{subsec:smooth}

As mentioned above, the performance of Hamiltonian amplification can be improved through the use of smooth continuous controls \cite{arenz2020amplification}. Along similar lines, one can amplify beamsplitter interactions by applying smooth global parametric controls to both interacting modes. This is achieved by the time-dependent Hamiltonian $ H_{\rm c} (t) = i f(t)(a^{2} - a^{\dagger 2}+ b^{2} - b^{\dagger 2})$ where the drive $f(t)$ is modulated according to $f(t) = \frac{\pi K }{ 2T_{\rm c}} \cos( \frac{2 \pi t }{T_{\rm c}})$. In this case, the dynamics are governed by the total time-dependent Hamiltonian
\begin{equation}\label{eq:H_tot(t)}
H_{\text{tot}}(t) = H_{\text{bs}} + H_{\rm c}(t)\,.
\end{equation}
For multiples of the drive period, $t = N T_{\rm c}$, the resulting evolution corresponds to the unitary transformation $U(NT_{\rm c}) = \exp(-i\bar{H} t)$, where $\bar{H}$ is given by the  Magnus expansion \cite{lidar2013quantum}. For sufficiently small $T_{\rm c}$, the dynamics are approximately governed by the first order of the Magnus expansion, namely $\bar{H}^{(0)} = \frac{1}{T_{\rm c}}  \int^{T_c}_{0} dt\, U_{\rm c}^\dagger(t) H_{\text{bs}}  U_{\rm c}(t) $, where $U_{\rm c}(t) = \exp \left(F(t)(a^{2} - a^{\dagger 2}+ b^{2} - b^{\dagger 2})\right)$ and $F(t) = \int^{t}_{0}ds\, f(s)$ is the integrated parametric drive. Within this framework, the amplified Hamiltonian is given by $\bar{H}^{(0)} = \lambda_{2} H_{\text{bs}}$, where $\lambda_{2} = I_{0}(K)$ with $I_{n}(K)$ the modified Bessel function of the first kind. The constant $K$ appearing in the amplitude of the control Hamiltonian therefore determines the magnitude of the speed-up. 

The amplification of the cross-Kerr interaction can also be obtained when both modes are continuously driven. This leads to an enhancement of $\lambda_{2} = I^{2}_0(K)+
2 \sum_{n=1}^{\infty} (-1)^n I_{4n}(K)\, I_{2n}(K)$ as discussed in Appendix \ref{sec:appendix1}. 

Despite the advantage in convergence of the continuous drives, a bang-bang implementation facilitates both the theoretical analysis and the numerical implementation. For this reason, in the present work we will restrict to bang-bang protocols as in \eqref{eq:two mode HA sequence beamsplitter} and \eqref{eq:two mode HA sequence cross-Kerr}.

\section{Amplification in the presence of noise and decoherence}\label{sec:AmpInPresenceofNoise}
\subsection{Noise models}\label{subsec:noisemodels}
In realistic settings, the dynamics governed by the system Hamiltonian are always accompanied by noise and decoherence arising from undesired fluctuations and couplings with the environment. In what follows, we will focus on different decoherence models that are relevant in experimental scenarios such as trapped ions and photonic systems \cite{RevModPhys.87.1419,Walls-Milburn}.

In the first model, we account for stochastic field fluctuations through random phase-space displacements. We model them by adding to the original system Hamiltonian $H_0$ a stochastic contribution corresponding to a generalized force described by 
\begin{equation}\label{eq:noisemodel_displacements}
H(t) =  H_{0} + \left[ i\, \boldsymbol{\alpha}(t) \cdot \boldsymbol{\xi}^{\dagger} + {\rm H. c.} \right], 
\end{equation}
where for compactness we use the notation $\boldsymbol{\alpha}\cdot\boldsymbol{\xi}^\dagger = \alpha_a\, a^\dagger + \alpha_b\, b^\dagger$. The displacement parameters $\alpha_a(t), \alpha_b(t)$ are complex uncorrelated variables whose time dependence can be chosen to represent different kinds of environmental fluctuations, with white noise and static disorder as limit cases. 

Alternatively, we also consider dissipative and/or dephasing processes originating from system-bath couplings described by a Lindblad master equation of the form \cite{gardiner2004quantum}
\begin{equation}\label{eq:lindbladian}
\pdv{\rho}{t} = \mathcal{L}(\rho) = -i [H_{0},\rho] + \sum_{L}  \eta_{L}   \mathcal{D}_{L}(\rho),
\end{equation}
where $\mathcal{D}_{L}(\cdot)= \left[ L (\cdot) L^{\dagger} - \frac{1}{2} \left( L^{\dagger}L(\cdot) +  (\cdot)L^{\dagger}L  \right ) \right]$ is the Lindblad superoperator associated with the jump operator $L$ and $\eta_{L}$ determines the respective dissipation rate. In particular, we will consider jump operators $L\in \{a,b\}$ to model losses due to the exchange of excitations with a bath in its vacuum state. Baths at non-zero temperatures can be described by adding also raising jump operators $L \in \{a^{\dagger}, b^{\dagger}\}$ acting at a lower rate than the corresponding lowering jump operators. Dephasing processes result from jump operators $L \in \{a^{\dagger}a,b^{\dagger}b\}$. Non-zero temperature and dephasing environments are discussed in Appendix \ref{sec:appendix3}. 

In order to assess the performance of Hamiltonian amplification in the presence of noise and decoherence, we implement the protocols in \eqref{eq:two mode HA sequence beamsplitter} and \eqref{eq:two mode HA sequence cross-Kerr} for the noise models given in \eqref{eq:noisemodel_displacements} and \eqref{eq:lindbladian}. That is, we replace the free evolution in the Hamiltonian amplification sequence by the evolution governed by either \eqref{eq:noisemodel_displacements} or \eqref{eq:lindbladian}. We neglect the effect of noise and decoherence during the bang-bang squeezing operations, an approximation that is consistent with the assumption that these are much faster than the free evolution steps.

\subsection{Results}

In the following, we study the fidelity of the preparation of Bell-type states through the amplification sequences \eqref{eq:two mode HA sequence beamsplitter}  and \eqref{eq:two mode HA sequence cross-Kerr} corresponding to the beamsplitter and cross-Kerr interaction, respectively, in the presence of noise and decoherence as described in \ref{subsec:noisemodels}. We will quantify the error by the infidelity, namely the quantity $1-F$ with $F$ being the fidelity between the target state and the state reached by the protocol applied.

First, we analyze the infidelity of the entangled state resulting from random displacements as described by Hamiltonian \eqref{eq:noisemodel_displacements}. Let us begin by considering the simple case of static disorder, where we assume $\alpha(t)$ and $\beta(t)$ are time-independent c-numbers that fluctuate from one realization to the next. When the Hamiltonian \eqref{eq:noisemodel_displacements} is enhanced by squeezing both modes, for $N\to \infty$ the resulting unitary is generated by a total Hamiltonian where the desired interaction strength is amplified by $\lambda_2$, which is given by $\cosh(2r)$ or $\cosh^2(2r)$ depending on the coupling Hamiltonian, whereas $\alpha$ and $\beta$ are amplified by $\cosh(r)$. Thus, due to the reduction in the total time required, the detrimental displacements are suppressed by a factor $\cosh(r)/\lambda_2$. As the squeezing strength is increased with $r\gg 1$, the suppression factor scales as $\propto e^{-r}$ for $H_{0} = H_{\text{bs}}$ and $\propto e^{-3r}$ for $H_{0} = H_{\text{cK}}$. 

A similar favorable scenario is expected for the more general case when $\alpha(t)$ and $\beta(t)$ fluctuate dynamically, so that a noise-tolerant speed up can still be obtained. This situation is analyzed numerically. In our simulations, for practical reasons we model time variations in the following way: for each time step $\Delta t$ we sample the parameters $\alpha$ and $\beta$ from a Gaussian distribution $\mathcal{N}(0,\sigma^2)$, so that our displacement noise is constant during each free evolution step but keeps no correlations from one step to the next. 

\begin{figure}[ht!]
 \centering
\includegraphics[scale = 0.675]{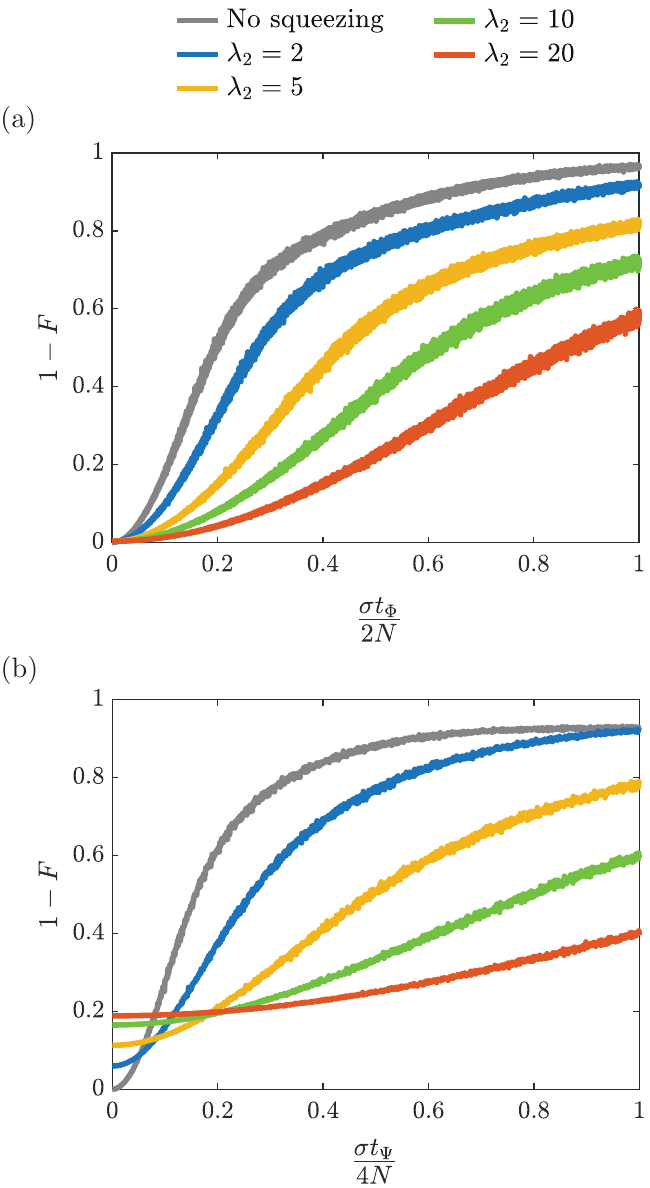}
\caption{Average infidelity in the preparation of (a) the entangled state \eqref{eq:entangledstate1} via the beamsplitter interaction \eqref{eq:beamsplitter} and (b) the entangled state \eqref{eq:entangledstate2} through the cross-Kerr interaction \eqref{eq:cross-Kerr}. Infidelities are plotted as a function of the noise strength $\sigma \Delta t$, with noise modeled as random displacements as described in \eqref{eq:noisemodel_displacements}. Here $\sigma$ is the standard deviation of the Gaussian distribution $\mathcal{N}(0,\sigma^2)$ from which displacement amplitudes are sampled, and $\Delta t$ is a fixed reference time step given by $\frac{t_{\Phi}}{2N}$ in (a) and $\frac{t_{\Psi}}{4N}$ in (b). The squeezing strength is choosen to achieve the amplification factors $\lambda_{2} = 2$ (blue), $\lambda_{2} = 5$ (yellow), $\lambda_{2} = 10$ (green), $\lambda_{2} = 20$ (red). For comparison, the averaged infidelity when no squeezing is present is shown in grey. In all cases, we set the number of Trotter steps to $N = 5$ in the squeezing sequences employed to amplify the desired interactions. The results were averaged over $10^3$ samples.
}
\label{fig:2std(1-F)}
\end{figure}

\begin{figure*}[ht!]
    \centering \includegraphics[width=0.825\textwidth]{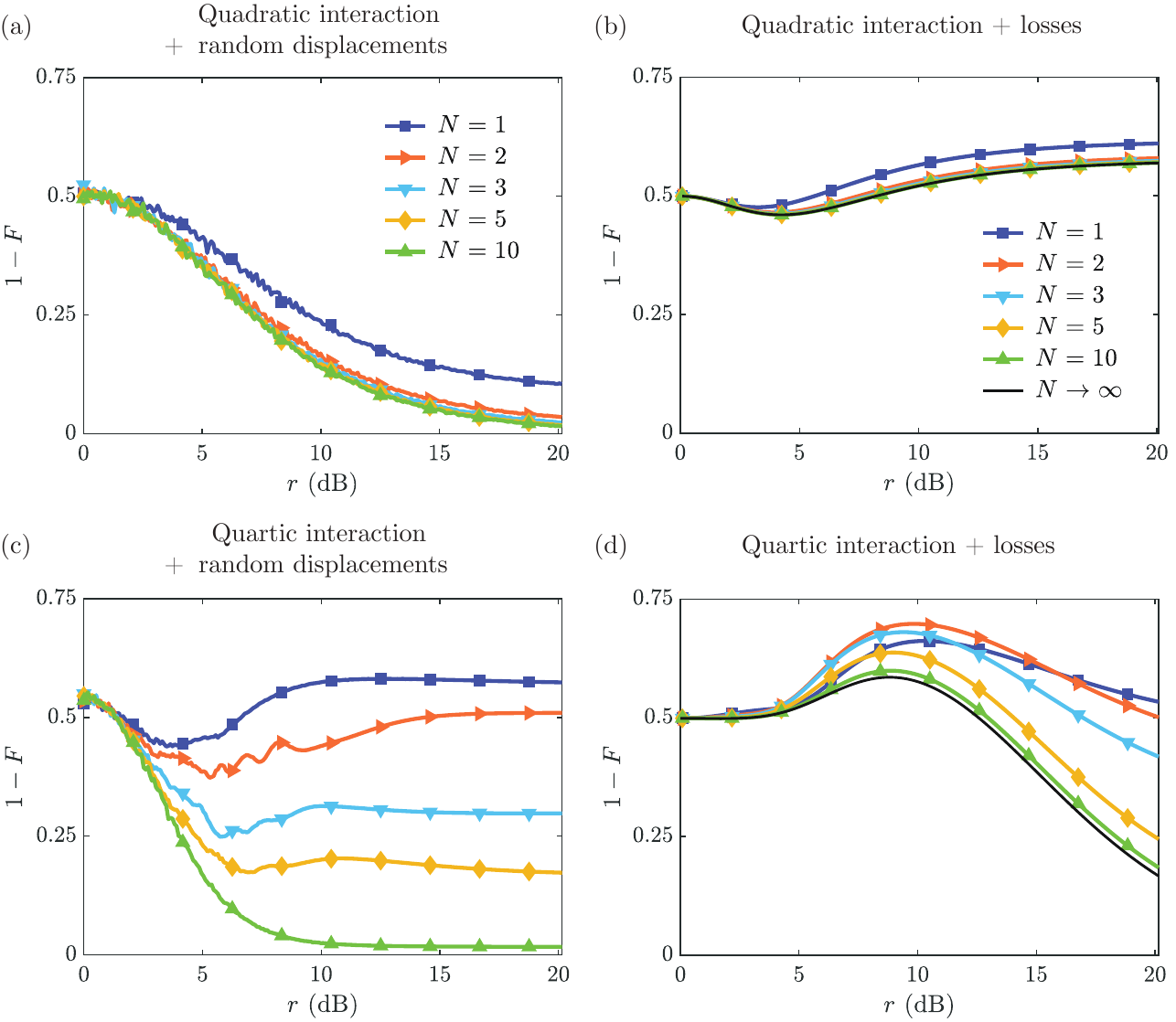}
    \caption{Infidelity in the generation of the entangled states \eqref{eq:entangledstate1} and \eqref{eq:entangledstate2}
    via the squeezing sequences \eqref{eq:two mode HA sequence beamsplitter} and \eqref{eq:two mode HA sequence cross-Kerr} respectively, as a function of squeezing strength employed in the amplification protocol.  Noise is modeled as random displacements in (a) and (c) and as excitation losses in (b) and (d). Subplots (a) and (b) correspond to beamsplitter interactions, while (c) and (d) were calculated for cross-Kerr couplings. The number of amplification steps is $N = 1$ (violet), $N= 2$ (orange), $N= 3$ (blue), $N= 5$ (yellow) and $N= 10$ (green). In all cases, we choose noise parameters such that in the absence of amplification, the infidelity is $\sim 0.5$. In subplots (a) and (c), the results were averaged over $10^3$ samples; for the case of losses, the evolution was modeled as a Lindblad equation (more details are given in Sec.~\ref{subsec:noisemodels}). The black line in (b) and (d) corresponds to taking the limit $N\to\infty$ in the expression for the full time evolution.
    }
    \label{fig:3r(dB)_1-F}
\end{figure*}

In Fig. \ref{fig:2std(1-F)} (a) we plot the average infidelity $1-F$ for the generation of the state $\ket{\Phi}$ via the amplified beamsplitter evolution obtained through the sequence \eqref{eq:two mode HA sequence beamsplitter} in the presence of displacement noise, as a function of the noise strength quantified by $\sigma \frac{t_\Phi}{2N}$, a representative measure of the displacement induced by the noise over a fixed time step. In the numerical simulations, we set $N = 5$ in the amplification sequence. We choose the squeezing strengths to achieve amplification factors of $\lambda_2 = 2$ (blue), $\lambda_2 = 5$ (yellow), $\lambda_2 = 10$ (green), $\lambda_2 = 20$ (red), and compare the infidelities with the case when no Hamiltonian amplification ($r = 0$) is used (grey). For each of the simulations, the duration of the time step is adjusted by the amplification factor, so that $\Delta t = \frac{t_\Phi}{2N\lambda_2}$. The infidelities were averaged over $10^3$ repetitions. The results show that, for moderate noise strengths, by squeezing up to $r \approx 16.02\,\text{dB}$ ($\lambda_2 = 20$), the infidelity can be reduced by an order of magnitude. Thus, within this noise model, Hamiltonian amplification successfully enhances the desired interaction while parametrically outpacing the effect of noise.

In Fig. \ref{fig:2std(1-F)} (b), we repeat the same comparison for the preparation of the state $\ket{\Psi}$ using a cross-Kerr interaction via the sequence \eqref{eq:two mode HA sequence cross-Kerr}. To facilitate the comparison with the previous case, once more we set $N = 5$ and average results over $10^3$ trajectories, while we adjust the time step as $\Delta t = \frac{t_\Psi}{4N\lambda_2}$. The results also indicate a significant reduction of the infidelities for a large region of parameters through the use of the amplification protocol. However, the figure shows that for very weak noise, the amplification sequence actually worsens the results. This is not a fundamental limitation but rather the consequence of the Trotter error, which becomes larger as the squeezing strength is increased. This problem can be counteracted by appropriately increasing the number of Trotter steps $N$ (see Appendix \ref{sec:appendix-TrotterError} for more details) \cite{arenz2020amplification, Tiwari:25}. 

We note that the results shown in Fig. \ref{fig:2std(1-F)}~(a) as well as those in Fig.~\ref{fig:3r(dB)_1-F}~(a) were computed through the symplectic representation of the dynamics, which we discuss in Appendix \ref{sec:appendix2}. The remaining cases were calculated with a truncated representation of the states in the Fock basis, checking for convergence of the results upon increase of the Hilbert space.

The interplay between squeezing strength and number of Trotter steps $N$  is investigated in Fig.~\ref{fig:3r(dB)_1-F}, keeping the noise strength fixed. More precisely, in Fig.~\ref{fig:3r(dB)_1-F} (a) to (d), we plot the infidelity for the generation of entangled states through beamsplitter (top row) and cross-Kerr interactions (bottom row). The noise model corresponds to random displacements (left column) and to excitation losses modeled through Lindblad operators (right column). We take noise parameters such that for no squeezing, the infidelity is $1-F\sim0.5$. The number of steps in the amplification protocol is $N = 1$ (violet), $N = 2$ (orange), $N = 3$ (blue), $N = 5$ (yellow), and $N = 10$ (green). As we increase the number of steps from $N = 1$ to $N = 10$, the Trotter error of the Hamiltonian amplification protocol generally decreases with $N$. However, the convergence can be very different depending on the noise model, the interaction considered and the squeezing strength. In general, for small values of $r$ the gain in increasing $N$ is very moderate; in contrast, a larger number of time steps can provide a significant improvement in the performance for sufficiently large squeezing, particularly in the case of cross-Kerr interactions. 

\subsection{Discussion}

The plots in Fig.~\ref{fig:3r(dB)_1-F} show an important contrast between scenarios where Hamiltonian amplification can, for the right parameter choices, successfully overcome noise and situations where this is not the case. Some intuition can be gained about this difference considering how the desired and undesired processes scale under the ideal amplification protocol, i.e., in the Trotter limit $N\to\infty$. Let us begin with the displacement noise corresponding to subplots Fig. \ref{fig:3r(dB)_1-F} (a) and (c). As discussed above, for the case of random displacements, one expects in the regime of large $r$ a suppression of noise scaling as $e^{-r}$ for beamsplitter interactions and $e^{-3r}$ for cross-Kerr couplings. Hamiltonian amplification is in this case in principle able to outperform displacement noise for these type of interactions, as confirmed by the numerical results. 

The case of excitation losses is more subtle because the amplification protocol does not induce a simple scaling factor of the Lindblad master equation, but also changes its form.
In the limit $N\to \infty$, the squeezing sequences that implement Hamiltonian amplification lead to an evolution governed by the superoperator
\begin{equation}
\label{eq:amplifiedLindbladiandecoherence_Ntoinfty}
\begin{aligned}
\mathcal{L}_{r}(\cdot) =& -i\,  [H_{\lambda_2}, \cdot]\\
&+\sum_{L}  \eta_{L} \left(\cosh^2(r)   \mathcal{D}_{L}(\cdot) + \sinh^2(r) \mathcal{D}_{L^\dagger}(\cdot)\right).
\end{aligned}
\end{equation}
Thus, the amplification protocol enhances the Hamiltonian evolution by a factor $\lambda_2$ while it amplifies the original Lindblad superoperator $\mathcal{D}_{L}$ by $\cosh^2(r)$ and introduces an additional Lindblad operator $\mathcal{D}_{L^\dagger}$ with a prefactor $\sinh^2(r)$. This new process corresponds to an effective bath that is no longer in a vacuum state, but rather in a thermal state, with a number of thermal excitations proportional to $\sinh^2(r)$. This allows us to identify an $r$-dependent effective bath temperature
 \begin{equation}
T(r) =  \frac{\omega}{k_{B} \ln \left[ \coth^{2}(r) \right]},
\end{equation}
where $k_{B}$ is the Boltzmann constant. 

For the case of beamsplitter interactions, the ratios between undesired and desired processes scale as $\frac{\cosh^2(r)}{\cosh(2r)}$ and $\frac{\sinh^2(r)}{\cosh(2r)}$ for losses and heating respectively, showing that no significant advantage should be expected from Hamiltonian amplification. This explains the poor results in Fig.~\ref{fig:3r(dB)_1-F}~(b). In contrast, for cross-Kerr interactions, the scaling is favorable: the undesired processes scale like $\frac{\cosh^2(r)}{\cosh^2(2r)}$ and $\frac{\sinh^2(r)}{\cosh^2(2r)}$. For sufficiently large squeezing strengths, these factors behave like $e^{-2r}$. This coincides with the decrease of the infidelity seen in Fig.~\ref{fig:3r(dB)_1-F}~(d) for large values of $r$ and $N$.

The plots in Fig.~\ref{fig:3r(dB)_1-F} also show that the behavior of the gate error with $r$ is generally non-monotonic. For small numbers of Trotter steps, a higher value of $r$ can lead to a more significant residual deformation of the state, associated with the gate error introduced by the protocol itself. This Trotter error depends on the system’s free evolution because it results from commutators of the control and free Hamiltonians. The ideal amplification protocol corresponds to $N$ approaching infinity with the value of $r$ fixed. In practice, one must find a balance of achievable values of $r$ and $N$ that lead to a sufficiently small gate error for the evolution of interest and the relevant noise model.

In summary, even in the most demanding conditions of large squeezing and many time steps, it is possible that Hamiltonian amplification fails to overcome noise, or it can even intensify it more than the desired interactions. As a rule of thumb, the amplification protocol will enhance more those processes that are of higher order in the number of creation and/or annihilation operators. Nonetheless, numerical analysis is required to decide if an advantageous regime can be reached in practice. At this point, experimental limitations will play a relevant role: constraints on how fast the squeezing protocol can be applied, how large $r$ can be made, and timing or amplitude errors in the control pulses will set restrictions on the achievable improvements \cite{arenz2020amplification}.

\section{Potential experimental implementations}

\label{sec:platforms}

We now turn to discuss the experimental feasibility of the Hamiltonian amplification protocol in different platforms. As mentioned above, proof-of-principle demonstrations of Hamiltonian amplification and other squeezing-based amplification protocols have been implemented in trapped ions \cite{burd2023experimental, doi:10.1126/science.aaw2884, scBurdQuantumampbosonic, PhysRevA.107.032425}, superconducting circuits \cite{PRXQuantum.5.020306} and photonic platforms~\cite{3bdr-njhy}. A similar technique based on trotterization leading to a transformed amplified Hamiltonian has been demonstrated in \cite{cai2025quantum} to enhance squeezing in a superconducting microwave cavity. Squeezing-based protocols have also been proposed to amplify interactions in optomechanics \cite{PhysRevLett.114.093602}, spin-phonon systems \cite{PhysRevLett.120.093601}, and atoms in optical cavities \cite{PhysRevLett.125.153602}. Although no experimental demonstrations of squeezing-enhanced interactions have been achieved in these last platforms up to date to the best of our knowledge, they are potential candidates for implementation, as discussed in the corresponding references above. 

We now briefly comment on some of the most promising experimental setups for the implementation of Hamiltonian amplification. In the prominent case of trapped ions, squeezing-based protocols for the speed-up of interactions have been demonstrated for single-mode displacements~\cite{doi:10.1126/science.aaw2884}, phonon-mediated qubit-qubit interactions \cite{scBurdQuantumampbosonic}, and spin-phonon couplings \cite{burd2023experimental, PhysRevA.107.032425}. Regarding the applicability of the proposal to the case of two-mode interactions, we note that 
several kinds of motional interactions have been implemented. For instance, coherent exchange of excitations in the motion of spatially separated groups of ions was demonstrated in \cite{Harlander_antenna}, while \cite{Hou_NaturePhys_2024} and \cite{chen2023scalable} showed beamsplitter operations between different normal modes of an ion chain. A Kerr coupling arising from nonlinearities in the Coulomb potential has been used for phonon counting in \cite{PhysRevLett.119.193602}. While two-mode couplings are generally weak, that weakness actually makes Hamiltonian amplification even more promising to implement in such systems. 

More in general, platforms that enable the generation of the desired bosonic interactions and sufficiently strong single-mode squeezing while maintaining low decoherence rates are good candidates for the experimental realization of the proposed noise-tolerant amplification. In particular, the generation of cross-Kerr interactions remains a central challenge across most physical platforms \cite{Tiwari:25, cai2025quantum}; consequently, systems in which cross-Kerr nonlinearities have been experimentally observed are especially relevant for the proposed scheme. In integrated photonic circuits, both beamsplitter \cite{hu2021chip,PhysRevResearch.7.023108} and cross-Kerr interactions \cite{feng2019enhanced, cui2022situ} as well as squeezing up to $7.8 \,\text{dB}$ \cite{ulanov2025quadrature} have been reported.  Notably, cross-Kerr modulated phase shifts as large as $0.3\,\text{mrad}$ \cite{venkataraman2013phase} have been realized in Rb-filled optical fibers. Ultra–low-loss integrated photonics \cite{tan2024ultra, bose2024anneal} significantly improve the prospects for an experimental realization of the proposed scheme within this architecture.

Finally, both kinds of bosonic interactions  \cite{kounalakis2018tuneable, liu2017circuit, copetudo2026direct, PhysRevA.72.064305} and squeezing up to $\approx 8 \,\text{dB}$ \cite{PRXQuantum.2.020323} have been reported in superconducting circuits. Recent advances in these platforms indicate a significant reduction in the characteristic decoherence rates \cite{PRXQuantum.4.030336,ganjam2024surpassing}. From these results, we expect a noise-tolerant speed-up in two-mode coupling could be attained with the largest squeezing strengths presently available in these systems.

In summary, although there remains an important gap between the parameters in our simulations and current state-of-the-art experiments, there is a multiplicity of platforms for which our results may become relevant in the future. Our models aim to establish scalings in order to better assess the potential advantages of the application of Hamiltonian amplification strategies. We note, however, that the actual experimental implementation of the protocols studied still poses significant challenges.

\section{Conclusion} \label{sec:conclusion}

A high-frequency parametric control that modulates the direction along which a quantum harmonic oscillator is squeezed can be utilized to amplify a broad range of bosonic interactions. However, the corresponding squeezing sequences not only amplify the desired processes but typically also enhance detrimental, uncontrolled dynamics. In this work we discussed when, for appropriate parameter choices, the desired unitary evolutions can be accelerated more than noise and decoherence. As illustrative cases of interest, we considered beamsplitter- and cross-Kerr interactions, corresponding to bosonic couplings that are quadratic or quartic in the annihilation and creation operators. 

In particular, we showed that noise modelled through random displacements can be outperformed in both cases; in these scenarios, Hamiltonian amplification provides not only a speed-up but also a decrease in the errors. 
We further showed that decoherence processes described as single-excitation losses can be overcome when cross-Kerr interactions are amplified, which can be useful to speed up the implementation of a two-qubit gate in photonic quantum computing \cite{Tiwari:25}. While in this case additional heating processes are introduced, the effective Lindblad operators that describe the amplified heating and loss processes are scaled up less than the cross-Kerr interaction.

Based on these findings, we discussed when the fidelity for preparing a Bell-type state by means of beamsplitter or cross-Kerr interactions can be significantly improved through modulated squeezing in the presence of noise and decoherence. In general, Hamiltonian amplification can be expected to scale favorably as long as the interaction to be amplified involves bosonic operators of higher order than the detrimental processes. For instance, for cross-Kerr interactions, pure dephasing will be amplified with a similar rate as the target Hamiltonian and therefore cannot be outperformed through the presented protocols. Nonetheless, even in cases when the scaling is convenient, numerical analysis is necessary to determine the parameter regime in which Hamiltonian amplification will perform well.

The amplification of desired processes through parametric control in noisy scenarios enables new ways to control open quantum systems. Rather than suppressing unwanted dynamics through control strategies such as dynamical decoupling \cite{Arenz_2017}, a fast modulation of the squeezing direction can result in a speed-up that overcomes noise and decoherence. We expect that a combination of both types of protocols, i.e., dynamical decoupling and Hamiltonian amplification, can be particularly powerful in hybrid systems where bosonic degrees of freedom couple to spin degrees of freedom  \cite{PhysRevX.15.021073, Hogg2024enhanced} to selectively enhance desired processes while detrimental ones are simultaneously suppressed. 

\bigskip

\section*{Acknowledgments}

C.C. acknowledges funding from CSIC and PEDECIBA (Uruguay). CA, and AT acknowledge funding
support from the Air Force Office of Scientific Research (AFOSR) under the award FA9550-24-1-0139.

\bibliography{Reference}

\appendix 
\section{Hamiltonian amplification of cross-Kerr interactions via continuous parametric drives}\label{sec:appendix1}
The cross-Kerr interaction can also be amplified through a parametric drive described by the Hamiltonian $ H_{\rm c} (t) = i f_{a}(t)(a^{2} - a^{\dagger 2})+ i f_{b}(t)(b^{2} - b^{\dagger 2})$ where $f_{a}(t) = \frac{\pi K }{4T_{\rm c}} \cos( \frac{ \pi t }{T_{\rm c}})$ and $f_{b}(t) = \frac{\pi K }{2T_{\rm c}} \cos( \frac{2 \pi t }{T_{\rm c}})$. In this case, the system is governed by the time-dependent Hamiltonian
$H_{\text{tot}}(t) = H_{\rm{cK}} + H_{\rm{c}}(t)$. For multiples of the drive period, $t = 2N T_{\rm c}$, and for sufficiently small $T_{\rm c}$, the dynamics of the system can be approximated as $U(2NT_{\rm c}) \approx \exp(-i\bar{H}^{(0)} t)$. Here $\bar{H}^{(0)}$ is the first order of the Magnus expansion given by $\bar{H}^{(0)} = \frac{1}{2T_{\rm c}}  \int^{2T_c}_{0} dt\, U_{\rm c}(t) H_{\text{cK}}  U_{\rm c}(t) $, where $U_{\rm c}(t) = \exp \left( F_{a}(t)(a^{2} - a^{\dagger 2})+ F_{b}(t)(b^{2} - b^{\dagger 2})\right)$ and $F_{a}(t) =  \int^{t}_{0}ds\, f_{a}(s)$ and $F_{b}(t) = \int^{t}_{0}ds\, f_{b}(s)$. As a result, the cross-Kerr interaction is amplified by the factor
\begin{equation}
\lambda_{2} = I^{2}_0(K)+
2 \sum_{n=1}^{\infty} (-1)^n I_{4n}(K)\, I_{2n}(K),    
\end{equation}
where  $I_{n}(K)$ is the modified Bessel function of the first kind. The global phase-space rotation $R(t)$ is governed by the Hamiltonian $H_R = \chi \,z(K)\, [a^{\dagger}a + b^{\dagger}b ]$ where
\begin{equation}
 z(K) =  I^{2}_0(K)-I_0(K)
+ 2 \sum_{n=1}^{\infty}(-1)^n I_{4n}(K)\,I_{2n}(K).
\end{equation}

\section{Symplectic representation of Hamiltonian amplification in the presence of random displacements}\label{sec:appendix2}

The beamsplitter interaction described by the Hamiltonian \eqref{eq:beamsplitter} can be alternatively represented as
\begin{equation}
    H_{\rm{bs}} = g (q_{a}q_{b}+p_{a}p_{b}), 
\end{equation}
where $q_{a},\,p_{a}$ and $q_{b},\,p_{b}$ are the dimensionless canonical momentum and position operators associated with the interacting modes $a$ and $b$. In the Heisenberg picture, the operators $q$ and $p$ transform under the squeezing transformations $S_{0, \pi} $ as ${S}^{\dagger}_{0,\pi}q{S}_{0,\pi}=e^{\mp r}q$ and $
{S}^{\dagger}_{0, \pi}p{S}_{0, \pi}=e^{\pm r}p$, whereas the unitary $U(\Delta t) = \exp(-iH_{\rm{bs}} \Delta t)$ gives 
\begin{equation}
\begin{aligned}
q_{a,b}(\Delta t) &= q_{a,b}(0) \cos(g \Delta t) + p_{b,a}(0) \sin(g \Delta t)\\
p_{b,a}(\Delta t) &= - q_{a,b}(0) \sin(g \Delta t) + p_{b,a}(0) \cos(g \Delta t).\notag
\end{aligned}
\end{equation}
By arranging the canonical operators in a vector $\textbf{x}(t) = \left( q_{a}(t),\, p_{b}(t),\, q_{b}(t),\, p_{a}(t)\right)^{T}$, we can obtain the operators modified by the squeezing sequence $U^{\text{bs}}_{N}(t)$ \eqref{eq:two mode HA sequence beamsplitter} through the relation \cite{serafini2023quantum}
\begin{equation}
    \textbf{x}(t) = \textbf{M}^N \textbf{x}(0),
\end{equation}
where
\begin{equation}
\textbf{M} = \begin{pmatrix}
    M & 0\\
    0 & M
    \end{pmatrix}\text{ , }M = 
\begin{pmatrix}
c^{2} - s^{2}e^{-4r} & 2cs\cosh(2r)\\
-2cs\cosh(2r) & c^{2} - s^{2} e^{4r}
\end{pmatrix},
\end{equation}
and we introduced the short-hand notation $c = \cos(g\Delta t)$ and $s = \sin(g\Delta t)$. We remark that the matrix $M$ is a symplectic matrix, i.e., it satisfies the relation $M^{T}\Omega M = \Omega $, where $\Omega = \begin{pmatrix}
    0 &1\\
    -1 & 0
\end{pmatrix}$ is the symplectic form. The matrix $\textbf{M}$ can be decomposed as $\textbf{M} = \textbf{A}\textbf{B}$, where  
\begin{equation}
\textbf{A} = 
\begin{pmatrix}
A & 0\\
0 & A
\end{pmatrix},~~~~
\textbf{B} = 
\begin{pmatrix}
B & 0\\
0 & B
\end{pmatrix},
\end{equation}
and
\begin{equation}
A = 
\begin{pmatrix}
c & e^{-2r}s\\
-e^{2r}s & c
\end{pmatrix},~~~~
B = 
\begin{pmatrix}
c & e^{2r}s\\
-e^{-2r}s & c
\end{pmatrix}.
\end{equation}

The non-degenerate $M$ can be decomposed in terms of mutually orthonormal left and right eigenvectors $v^{(l)}_{m}$ and $v^{(r)}_{m}$ as
\begin{equation}
M = \sum_{m=1,2} \lambda_{m}   v^{(r)}_{m}   v^{(l)}_{m},
\end{equation}
where the corresponding eigenvalues are given by
\begin{equation}
    \lambda_{m} = 1-2{\tilde s}^2 + i (-1)^{m+1} \, 2 \tilde s\sqrt{1-{\tilde s}^2},
\end{equation}
where $\tilde s = \sin(g\Delta t)\cosh(2r)$. The eigenvalues are real and negative for $\tilde s \geq 1$, whereas for $\tilde s < 1$ the eigenvalues are
\begin{equation}
    \lambda_{m} = e^{(-1)^{m+1}2i\tilde \phi},
\end{equation} 
where $\tilde \phi = \sin^{-1}(\tilde s)$. The corresponding left and right eigenvectors can be
written in terms of the matrix elements of $M$ as
\begin{equation}
     v_{m}^{(r)} = 
   \begin{pmatrix}
       M_{12} \\  \lambda_{m}- M_{11}  
   \end{pmatrix},\,
     v_{m}^{(l)} = Z_{m} 
   \begin{pmatrix}
       M_{12}, &  M_{11}-\lambda_{m}
   \end{pmatrix},
\end{equation}
where $Z_m = \frac{1}{M_{12}^2-(\lambda_m-M_{11})^2}$. We thus arrive at 
\begin{equation}
\begin{aligned}
    M^N &=\\
    &\sum_{m = 1,2} \lambda_m^N Z_m 
    \begin{pmatrix}
        M_{12}^2 & M_{12}(M_{11}-\lambda_m)\\
        -M_{12}(M_{11} -\lambda_m ) & -(\lambda_m -M_{11})^2
    \end{pmatrix}.
\end{aligned}
\end{equation}

One can describe the effective Hamiltonian which generates the unitary $U^{\text{bs}}_{N}(t)$ with the relation
\begin{equation}
e^{\mathbf{\Omega}\textbf{H} t} = \textbf{M}^N,
\end{equation}
where $\mathbf{\Omega} = \begin{pmatrix} 0 & \Omega \\ \Omega & 0 \end{pmatrix}$ and \textbf{H} is a symmetric matrix that takes the form
\begin{equation}
 \textbf{H} = 
 \begin{pmatrix}
  0 & \bar{H}\\
  \bar{H} & 0 
 \end{pmatrix}, 
 \bar{H} =
 \begin{pmatrix}
u & v\\
v & u
\end{pmatrix},
\end{equation}
where $u, v \in \mathbb{R}$  depend on the system parameters $r$, $N$, $g$ and $t$. The effective Hamiltonian $H_{\text{eff}} = \frac{1}{2} \textbf{x}^{T}\textbf{H}\,\textbf{x}$ is thus given by
\begin{align}\label{eq: effective Hamiltonian}
H_{\text{eff}} =& u \left( q_{a}q_{b} + p_{a}p_{b} \right) + \frac{v}{2}\left ( \{q_{a},p_{a}\} + \{q_{b},p_{b}\} \right).
\end{align}
We remark that for $N \to \infty$, the coefficients become $u \to \lambda g $ and $v \to 0$.

In total we obtain $\textbf{x}(t) = \textbf{M}^{N} \textbf{x}(0) + \boldsymbol{\mathcal D}$, where
\begin{equation}
\boldsymbol{\mathcal{D}} = \sum^{N}_{j = 1} \left( \textbf{M}^{N-j} \textbf{D}_{j,2} +
\textbf{M}^{N-j}\textbf{A}
\textbf{D}_{j,1}
\right),
\end{equation}
and the vector $\textbf{D}_{j,1}$ describes the net displacement of the canonical operators after the $S^{\dagger}_{\pi,\pi}$ operation in \eqref{eq:two mode HA sequence beamsplitter} at the $j$th Trotter step. It is explicitly given by
\begin{equation}
    \textbf{D}_{j,1} = \textbf{E}_{+} \int^{(2j-1)\Delta t}_{(2j-2)\Delta t} \begin{pmatrix}
   e^{g \Omega (\Delta t - s) }& 0\\
   0&e^{g \Omega (\Delta t - s) }
\end{pmatrix} \textbf{d}_{\alpha_{1},\beta_{1}}\, ds,
\end{equation}
where 
\begin{equation}
\textbf{E}_{+} = 
\begin{pmatrix}
E_{+} & 0\\
0  & E_{+}\\
\end{pmatrix}
\text{, }
{E}_{+} = 
\begin{pmatrix}
e^{r} & 0\\
0  & e^{-r}\\
\end{pmatrix},
\end{equation}
and $\textbf{d}_{\alpha_{1},\beta_{1}} = \sqrt{2} \left(\text{Re}(\alpha_{1}), \text{Im}(\beta_{1}), \text{Re}(\beta_{1}) , \text{Im}(\alpha_{1}) \right)^{T}$. In a similar fashion, the vector $\textbf{D}_{j,2}$ describes the net displacement of the canonical operators after the $S^{\dagger}_{0,0}$ operation in \eqref{eq:two mode HA sequence beamsplitter} at the $j$th Trotter step given by
\begin{equation}
    \textbf{D}_{j,2} = \textbf{E}_{-} \int^{2j\Delta t}_{(2j-1)\Delta t} \begin{pmatrix}
   e^{g \Omega (\Delta t - s) }& 0\\
   0&e^{g \Omega (\Delta t - s) }
\end{pmatrix} \textbf{d}_{\alpha_{2},\beta_{2}}\, ds,
\end{equation}
where
\begin{equation}
\textbf{E}_{-} = 
\begin{pmatrix}
E_{-} & 0\\
0  & E_{-}\\
\end{pmatrix}
\text{, }
{E}_{-} = 
\begin{pmatrix}
e^{-r} & 0\\
0  & e^{r}\\
\end{pmatrix},    
\end{equation}
and $\textbf{d}_{\alpha_{2},\beta_{2}} = \sqrt{2} \left(\text{Re}(\alpha_{2}), \text{Im}(\beta_{2}), \text{Re}(\beta_{2}) , \text{Im}(\alpha_{2}) \right)^{T}$. 

Since random displacements shift the operators transformed by the matrix $\textbf{M}^{N}$, in the Schr\"{o}dinger picture the dynamics resulting from Hamiltonian amplification of the beamsplitter interaction can be described by a unitary governed by $H_{\rm{eff}}$ \eqref{eq: effective Hamiltonian}, followed by the displacement operators which correspond to the net displacement collected in the vector $\boldsymbol{\mathcal D}$. As such, the sequence of unitary operators that exactly describes Hamiltonian amplification in the presence of random displacements \eqref{eq:noisemodel_displacements} is of the form 
\begin{equation}
D(\gamma,\nu)e^{-iH_{\text{eff}}t},
\end{equation}
where $D(\gamma,\nu) = D(\gamma)_{a} D(\nu)_{b}$, and $D(\gamma)_{a}$ and $D(\nu)_{b}$ are the displacement operators acting on the $a$ and the $b$ modes, respectively. We used this expression to obtain the numerical simulation results shown in Fig. \ref{fig:2std(1-F)}~(a)
 and Fig. \ref{fig:3r(dB)_1-F}~(a).

\section{Trotter errors in Hamiltonian amplification of beamsplitter and cross-Kerr interactions}\label{sec:appendix-TrotterError}

In this section, we analyze the errors obtained due to insufficient Trotterization of the squeezing sequences employed in amplification of the beamsplitter and cross-Kerr interactions. In order to better identify the effect of the Trotter error, we include no noise in the simulations, and the dynamics are entirely unitary.

\begin{figure}[ht!]
 \centering
\includegraphics[scale = 0.675]{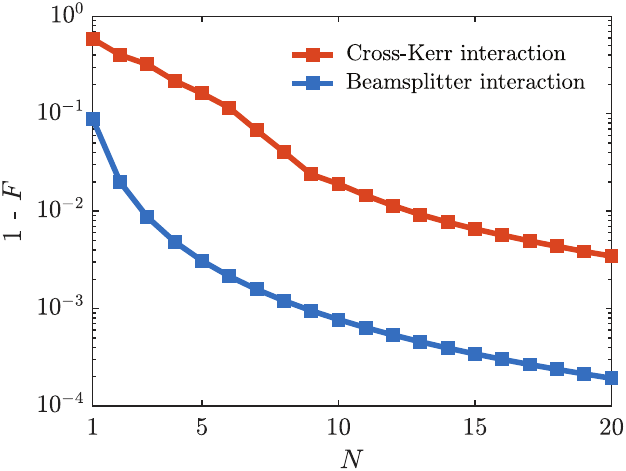}
\caption{Infidelity 
in the preparation of the entangled state \eqref{eq:entangledstate1} via the beamsplitter interaction \eqref{eq:beamsplitter} (blue) and the entangled state \eqref{eq:entangledstate2} through the cross-Kerr interaction \eqref{eq:cross-Kerr} (red), as a function of the number of Trotter steps employed in the squeezing sequences. In both cases, we set $r$ such that the desired interactions are amplified by $\lambda_{2} = 20$.
}
\label{fig:appndx_N(1-F)}
\end{figure}

In Fig. \ref{fig:appndx_N(1-F)} we plot the infidelity in the preparation of the entangled state \eqref{eq:entangledstate1} via the beamsplitter interaction (blue) and the entangled state \eqref{eq:entangledstate2} via the cross-Kerr interaction (red) as a function of the number of Trotter steps implemented in the squeezing sequences.
We use the same entangling times as in the numerical simulations shown in Fig. \ref{fig:2std(1-F)} and Fig.~\ref{fig:3r(dB)_1-F} of the main text. Similarly, the numerical results for beamsplitter interaction were obtained using the symplectic representation of the dynamics (see Appendix \ref{sec:appendix2}), whereas for cross-Kerr interaction, we utilize a truncated Fock-space approximation.

The results in the figure clearly indicate, as expected, a decrease in the Trotter error when the number of steps $N$ is increased. It is also apparent that the error depends strongly on the model, with cross-Kerr couplings leading to infidelities that can be more than one order of magnitude larger by than the ones obtained for beamsplitter interactions. This explains the very different behaviour observed in Fig.~\ref{fig:2std(1-F)} in the limit of weak noise. We note once more that, as discussed in \cite{arenz2020amplification}, the use of smooth controls can also improve performance by eliminating the second-order Trotter error.

\section{Hamiltonian amplification in the presence of dephasing or thermal baths}\label{sec:appendix3}

In the following, we discuss how a two-mode Hamiltonian amplification protocol of the form of \eqref{eq:two mode HA sequence beamsplitter} modifies the effect of dephasing and thermal baths. We restrict to the Trotter limit $N\to\infty$, which corresponds to the ideal amplification scenario in the absence of decoherence, and which is useful to gain intuition about the scaling of the processes involved with the squeezing strength $r$.

The effect of pure dephasing can be modeled by a Lindblad master equation of the form 
\begin{equation}\label{eq: master euqation for dephasing}
\begin{aligned}
\pdv{\rho}{t}  =& 
 -i\,  [H_{0}, \rho] - \frac{1}{2}  \sum_{L} \gamma_{L} [L^{\dagger}L,[L^{\dagger}L,\rho] ] 
\end{aligned}
\end{equation}
where $L \ \in \{ a ,b  \}$ and $\gamma_{L}$ is the dephasing rate. 
When Hamiltonian amplification is applied, in the Trotter limit the dynamics are governed by the superoperator 
\begin{equation}
\label{eq: dephasing channel}
\begin{aligned}
\mathcal L_{r}(\cdot) =& -i\,  [H_{\lambda_2}, \cdot]  - \frac{1}{2}  \cosh^2(2r)  \sum_{L } \gamma_{L} [L^{\dagger}L ,[L^{\dagger}L ,\cdot] ] \\
&+\frac{1}{4} \sinh^2(2r) \Biggl(\sum_{L} \gamma_{L} \left( \mathcal{D}_{L^2} (\cdot)  + \mathcal{D}_{L^{\dagger 2}} (\cdot) \right)\\
& - \frac{1}{2}  \sum_{L} \gamma_{L} \left( [L^2,[L^2,\cdot]] +   [L^{\dagger 2},[L^{\dagger 2},\cdot]] \right) \Biggr).
\end{aligned}
\end{equation}

Finally, we consider the effect of a thermal bath of non-zero temperature described by the master equation 
\begin{equation}
\begin{aligned}
\pdv{\rho}{t}  &= -i [H_{0}, \rho] + \sum_{L} \eta_{c,L} \,\mathcal{D}_{L}(\rho)+ \sum_{L^{\dagger} }  \eta_{h,L^{\dagger}}\, \mathcal  D_{L^{\dagger} }(\rho)
\end{aligned}
\end{equation}
where $\eta_{c,L}$ and $\eta_{h,L^{\dagger}}$ are, respectively, the loss and heating rates associated with each mode. In the Trotter limit, the dynamics are governed by the superoperator
\begin{equation}
\begin{aligned}
\mathcal{L}_{r}(\cdot) =& -i\,  [H_{\lambda_2}, \cdot]\\
&+\sum_{L}  \eta_{c,L} \left(\cosh^2(r)   \mathcal{D}_{L}(\cdot) + \sinh^2(r) \mathcal{D}_{L^\dagger}(\cdot)\right)\\
&+\sum_{L^{\dagger}}  \eta_{h, L^{\dagger}} \left(\cosh^2(r)   \mathcal{D}_{L^{\dagger}}(\cdot) + \sinh^2(r) \mathcal{D}_{L}(\cdot)\right).
\end{aligned}
\end{equation}

\end{document}